# A Novel Alluaudite-Type Vanadate, $Na_2Zn_2Fe(VO_4)_3$ : Synthesis, Crystal Structure, characterization and Magnetic Properties

Nour El Houda Lamsakhar [a,*], Mohammed Hadouchi[a,*], Mohammed Zriouil[a], Abderrazzak Assani[a], Mohamed Saadi[a], Abdelilah Lahmar[b], Mimoun El Marssi[b] and Lahcen El Ammari[a]

[a]Laboratoire de Chimie Appliquée des Matériaux, Centre des Sciences des Matériaux, Faculty of Sciences, Mohammed V University in Rabat, Avenue Ibn Battouta, BP 1014, Rabat, Morocco.

[b]Laboratoire de Physique de La Matière Condensée (LPMC), Université de Picardie Jules Verne, Amiens, France.

**Abstract**

A novel alluaudite-type vanadate, $Na_2Zn_2Fe(VO_4)_3$ has been successfully synthesized in its single crystal and polycrystalline forms. Its crystal structure was determined by means of X-ray diffraction measurements. This new vanadate crystallizes in monoclinic system with a space group *C2/c*. The crystal structure of this vanadate displays an open transition-metals based framework enclosing two kind of tunnels where the $Na^+$ cations are located. Moreover, the purity of the obtained powder was confirmed by X-ray powder diffraction analysis. Furthermore, this new vanadate was characterized by scanning electron microscopy, Infrared spectroscopy and by magnetic measurements. The magnetic susceptibility showed the antiferromagnetic behavior of this vanadate with Curie-Weiss temperature of $\theta_{cw} = -42.36$ K, consistently with the expected magnetic interactions between $Fe^{3+}$ cations.

*Corresponding authors :

Mohammed Hadouchi : hadouchimohammed@yahoo.com

Nour El Houda Lamsakhar : nourelhouda_lame@yahoo.fr

## Introduction

Transition metals-based materials with open frameworks attracts a great deal of interest because of their various potential applications such as effective electrode materials for rechargeable batteries [1,2] and interesting magnetic properties [3,4].

In this spirit, our team is contributing in the research and valorization of new metal transitions-based materials adopting open three-dimensional framework. Indeed, our attention is focused on the synthesis and characterization of new materials belonging to the well-known family, namely, "Alluaudite", of general formula $A(2)A(1)M(1)M(2)_2(XO_4)_3$, established by Moore (1971) [5]. The A sites may be occupied by mono- and / or divalent cations, while the M cationic sites correspond to an octahedral environment occupied by bi- or trivalent cations.

In the context of the exploration of new transition-metals based materials, alluaudite family [6] is a promising candidate due its high structural stability that can accommodate wide selection of transition metals. Indeed, several works have been dedicated to the synthesis of new alluaudite compounds due to their promising properties such as active electrode material for Na and/or Li-ion batteries[7,8]. As a contribution in this field, i.e, the exploration of new open framework transition metals-based phosphates and vanadates belonging to the well-known, alluaudite family, our previous investigations yielded to a new alluaudite-type phosphates, e.g., $Na_2Co_2Cr(PO_4)_3$ [9] and also vanadates, e.g., $A_2Co_2Fe(VO_4)_3$ (A = Ag, Na) [10,11]. However, the literature survey shows a very few of alluaudite-type vanadates that have been reported to date.

Herein, we report in this work, the synthesis, crystal structure and characterization of a new vanadate belonging to alluaudite class, i.e. $Na_2Zn_2Fe(VO_4)_3$. The magnetic properties of this vanadate were also discussed in this work.

## Experimental Section

**Single crystals synthesis.** The crystals of this novel vanadate were grown from a molten mixture of $NaNO_3$, $Zn(CH_3COO)_2.2H_2O$, $Fe(NO_3)_3.9H_2O$ and $V_2O_5$ in the molar ratio of Na : Zn : Fe : V = 2 : 2 : 1 : 3. The mixture was dissolved in distilled water with a few drops of $HNO_3$ and maintained under stirring. After the complete dissolution of the reactants, the solution was heated on a hot plate until dryness. The obtained powder was transferred to a platinum crucible and heated to the melting temperature of 1033 K, held at this temperature for one hour and followed by a slow rate cooling to room temperature, i.e., 5 Kh$^{-1}$. The resulting product contained a homogeneous parallelepipedic yellow crystals corresponding to the title compound.

**Powder synthesis.** The powder form of this new vanadate was obtained by conventional solid-state reaction. The same reactants used in single crystals synthesis were weighted in the same molar ratio and similar procedure was performed, i.e., from the dissolution of the reactants to the complete dryness of mixture. The obtained solid was transferred to furnace in a gold crucible. After a successive heat treatments at 473, 573, 673, 773 and finally at 853 K for 24h for each treatment with intermittent grinding, a yellow powder was obtained. The purity of the resulting powder was confirmed by powder X-ray diffraction analysis.

**Scanning electron microscopy.** The elemental analysis and the morphology of the single crystal and the powder were carried out by JEOL JSM-IT100 InTouchScope™ scanning electron microscope equipped with energy dispersive X-ray spectroscopy analyzer (EDS).

**Single crystal and powder X-ray diffraction.** A single crystal with suitable dimensions was carefully chosen for X-ray data collection at room temperature using a Bruker X8 Apex diffractometer equipped with an Apex II CCD detector and Graphite monochromator for Mo K$\alpha$ radiation, $\lambda$= 0.71073 Å. The software APEX2 was used for data collection and SAINT for cell refinement and data reduction. A total number of 35694 reflections were measured in the $\theta$ range of 2.4-35°. The crystal structure was solved using direct method and refined by SHELXT 2013[12] and SHELXL 2013[13] programs incorporated in the WinGX [14] program. Absorption corrections were made using the SADABS program [15] from equivalent reflections on the basis of multiscans. For structural drawing, DIAMOND [16] was used.

For confirming the formation of a single alluaudite phase in the synthesized powder, an X-ray powder diffraction pattern was recorded at room temperature using a Siemens D5000 powder diffractometer operating with $\theta$-$2\theta$ scan mode and CuK$\alpha$ radiation ($\lambda$ = 1.5406 Å). The pattern was measured over the $2\theta$ angle range of $10° \leq 2\theta \leq 70°$ with a step size of 0.04° and 30 s per step counting time.

**FT-Infrared spectroscopy.** FT-infrared spectrum of the powder sample (in KBr pellet) was measured by a Perkin Elmer RX-I model spectrometer with spectral resolution equal to 4 cm$^{-1}$ over the entire frequency range of 400–1400 cm$^{-1}$

**Magnetic measurements.** Magnetic measurements of Na$_2$Zn$_2$Fe(VO$_4$)$_3$ were performed by a Physical Property Measurement System (PPMS) DynaCool magnetometer on the powder sample sealed in a gelatin capsule. DC magnetization measurements were made between 300 and 2 K in zero field cooled (ZFC) mode with an applied field of 100 Oe. The ZFC mode was performed by cooling the sample from 300 down to 2 K in the absence of a magnetic field. Afterward, a magnetic field of 100 Oe was applied and the data were collected on heating the sample up to 300 K. Isothermal magnetization versus field was measured at different temperatures.

## Results and Discussion

**Scanning electron microscopy results.** SEM micrographs of the synthesized powder is shown in Fig.1. The EDS spectrum is presented in Fig. S1 (supporting information). As shown in Fig. 1, the powder is formed by a packed spherical grains with different micrometric sizes. The EDS analysis confirmed the existence of only Na, Zn, Fe, V and oxygen atoms with an atomic percentage very close the target composition

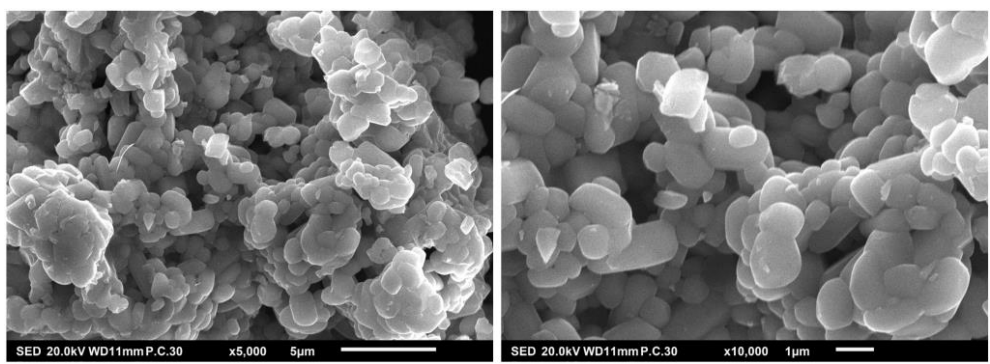

**Fig.1** SEM micrographs for $Na_2Zn_2Fe(VO_4)_3$

**Structural Description.** The novel vanadate, $Na_2Zn_2Fe(VO_4)_3$, belonging to the alluaudite family, crystallizes in the monoclinic system with space group *C2/c*. In this structure, the asymmetric unit consists of two Na atoms, two Zn atoms, one Fe atom and six oxygen atoms. Four of these atoms are located in special positions. i.e., Na1 is located in symmetry centre, special position 4b, while Na2, Zn2 and V2 atoms are located in twofold rotation axis sites 4e of the space group C2/c. The other remaining atoms are located in general position. In this vanadate the atoms Fe1 and Zn1 share the same site, 8f, with occupancy rate of 50 % each. Similar mixed sites were observed in alluaudite-type vanadates, i.e., $Na_2Co_2Fe(VO_4)_3$, [11] $Ag_2Zn_2Fe(VO_4)_3$[17] as well as in phosphates, i.e., $Na_2Co_2Cr(PO_4)_3$ [9], $Na_2Co_2Fe(PO_4)_3$ [18]. Bond valence sums calculations (Brown & Altermatt,) [19] for all atoms are in good agreement with the oxidation states (shown in supporting information). In the structure of $Na_2Zn_2Fe(VO_4)_3$, two edge-sharing [(Zn1,Fe1)$O_6$] octahedra, leads to the formation of [(Zn1,Fe1)$_2O_{10}$] dimers. These dimers are also linked by a common edge to [Zn2$O_6$] octahedra. The linkage of alternating [Zn2$O_6$] octahedra and [(Zn2,Fe2)$O_{10}$] dimers leads to infinite zigzag chains along the [10$\bar{1}$] direction (see Fig.S2). These chains are bonded together via vertices-sharing by V1O$_4$ tetrahedra in such a way as to build layers parallel to (010) plane (see Fig.S3). The 3D framework consists of the layers succession through [010] direction which opens two types of large tunnels, in which the Na+ cations are located as shown in Fig. 2. The Na1 atoms located in the first tunnel is bonded to eight oxygen atoms with Na1–O distances in the range of 2.4235 (15) and 2.8931 (16) Å, while Na2 in the second type of tunnels is surrounded by six oxygen atoms with Na2–O distances varying between 2.431 (2) and 2.827 (2) Å. Crystal data, data collection and structure refinement details are summarized in Table S1. Fractional atomic coordinates, atomic displacement parameters, geometric parameters and bond valence sums calculation are given in Tables S2, S3, S4 and S5 respectively (in supplementary material).

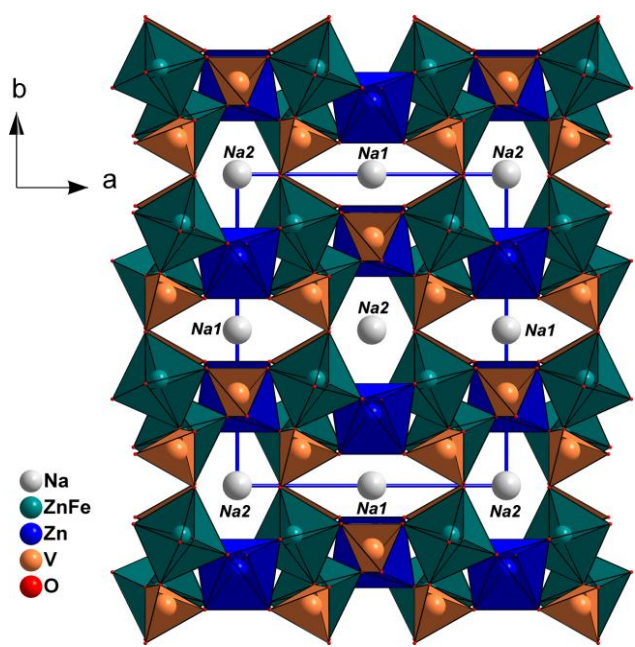

**Fig. 2** Polyhedral representation of the crystal structure of Na$_2$Zn$_2$Fe(VO$_4$)$_3$ showing the 3D framework which opens two type of tunnels running along the *c*-axis containing Na$^+$ cations

**Powder X-ray diffraction and Infrared spectroscopy.** The single phase of the prepared powder of Na$_2$Zn$_2$Fe(VO$_4$)$_3$ was confirmed by powder X-ray diffraction analysis. The comparison with the experimental pattern and the simulated one from single X-ray data confirmed the purity of the synthesized powder (Fig.3a). Moreover, the powder X-ray pattern was fitted by Le Bail refinement method [20] using JANA2006 software [21]. The refinement yielded to good fit between the experimental and the calculated patterns (see Fig. 3b), thus, confirming the pure phase of the synthesized powder. The obtained unit cell parameters are closer to those obtained from single crystal data. Le Bail refinement parameters are given in Table S6 (supporting information).

The room temperature Infrared spectrum measured for $Na_2Zn_2Fe(VO_4)_3$ powder is presented in Fig. S4 (supporting information), highlighting typical vibrational modes of $VO_4^{3-}$ groups. Indeed similar IR spectrum was reported in $Na_2Co_2Fe(VO_4)_3$ alluaudite vanadate [10]. The observed bands at 961, 850, 748 and at 685 cm$^{-1}$ are most likely assigned to $v_3$ antisymmetric stretching modes of $VO_4^{3-}$ [10,22], whereas, the band located at 476 cm$^{-1}$ is attributed to $v_4$ antisymmetric bending mode of $VO_4^{3-}$.

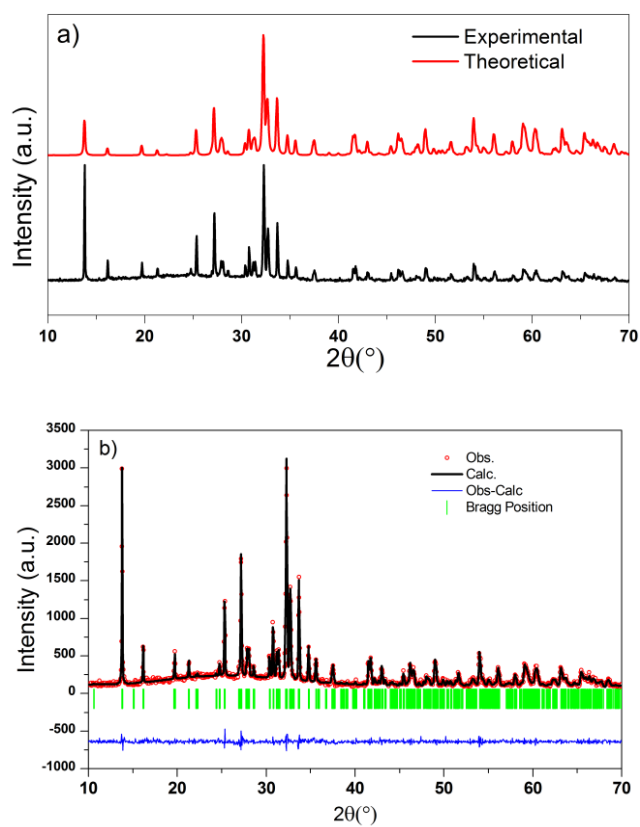

**Fig. 3** (a) Comparison between the experimental and the simulated X-ray patterns of $Na_2Zn_2Fe(VO_4)_3$. (b) Le Bail refinement of PXRD pattern.

**Magnetic Properties.** The magnetic susceptibility ($\chi = M/H$) measured at 100 Oe, within the temperature region of 2-300 K in ZFC condition is presented in Fig 4a. The data were corrected from diamagnetic signal (-252.10$^{-6}$ emu.mol$^{-1}$). The calculated inverse magnetic susceptibility $\chi^{-1}$ shows linear behavior above 50 K (Fig. 4b), thus, $\chi^{-1}$ versus $T$ follows well the Curie-Weiss law and the fitting gave a paramagnetic Curie-Weiss temperature $\theta_{cw}$ = − 42.36 K indicating the dominant interactions between $Fe^{3+}$ cations are antiferromagnetic. Indeed, closer Curie-Weiss constant values were reported in iron-based vanadates, e.g., $\theta_{cw}$ = − 46.7 K in $Mg_2FeV_3O_{11-x}$ [23], while a $\theta_{cw}$ = − 55.1 K was reported in its homolog

Zn$_2$FeV$_3$O$_{11}$[23]. Furthermore, a Curie-Weiss constant of $\theta_{cw}$ = − 30 K was observed in the iron vanadate FeVO$_4$ [24]. The resulting Curie constant $C$ = 4.63 emu K mol$^{-1}$ per formula unit, yielded to an effective magnetic moment per Fe$^{3+}$ of $\mu_{eff}$ = 6.09 $\mu_B$. This value agrees with high spin Fe$^{3+}$(S = 5/2, $\mu_s$ = 5.92 $\mu_B$). As described in the structural features of this vanadate, the framework is built-up from one type of zigzag chains running along [10$\bar{1}$] direction, which consist of the edge-sharing between ZnO$_6$ octahedra and the mixed-site dimers (Zn/Fe)$_2$O$_{10}$. The distance Fe$^{3+}$—Fe$^{3+}$ in the dimers is 3.21 Å and the angle Fe$^{3+}$—O1— Fe$^{3+}$ is 97.13 °(see Fig.5). The direct cation-anion-cation super-exchange interactions in this dimer, according to Goodenough - Kanamori - Anderson (GKA) semi-empirical rules [25–27], are antiferromagnetic consistently with the negative Curie-Weiss temperature. Another type of super-exchange interactions is expected between these chains in [001] direction (Fig.6), which involves the Fe$^{3+}$ cations in (Zn1/Fe1)O$_6$ octahedra through V(1)O$_4$ tetrahedra. These long-range magnetic interactions of a 180 ° cation-anion-anion-cation type which are weaker than the first type of interactions in the dimers, however, they are expected to be antiferromagnetic with reference to GKA rules. In the χ($T$) curve (Fig.4a), the transition from the paramagnetic state to the antiferromagnetic state could not be observed even at 2 K, this might be explained by the long-range ordering temperature is lower than 2 K. Similar χ($T$) curve was observed in SrMn$_3$P$_4$O$_{14}$ [28].

For further inspecting the magnetic behavior of this vanadate, *M-H* curves were measured between -10 kOe and 10 kOe at the temperatures of 2, 10, 20 and 50 K as presented in Fig. 7. At 2 K, the magnetization increases with increasing the field almost linearly and no open hysteresis was observed, thus confirming the presence of antiferromagnetic interactions causing compensation of Fe$^{3+}$ moments. Similar *M-H* curve was observed in Mg$_2$FeV$_3$O$_{11-x}$ [29] and in Pb$_2$FeV$_3$O$_{11}$ [30]. Moreover, the *M-H* curves at temperatures of 10, 20, 50 K have a perfect linear shape signifying the paramagnetic state. However, for deep investigation the magnetic data of this vanadate, theoretical magnetic study is needed, e.g., Monte Carlo simulation is a beneficial method for investigating the magnetic properties of magnetic materials [31–33], and will be the subject of our future work.

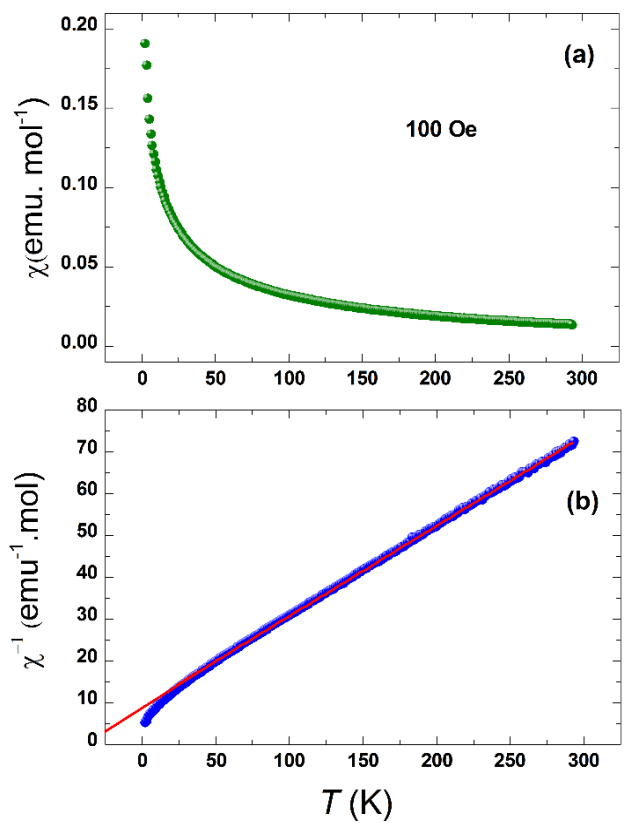

**Fig. 4** (a) Temperature dependence of magnetic susceptibility measured at 100 Oe in ZFC mode in the temperature range of 2-300 K. (b) The Curie-Weiss fitting of the reciprocal magnetic susceptibility

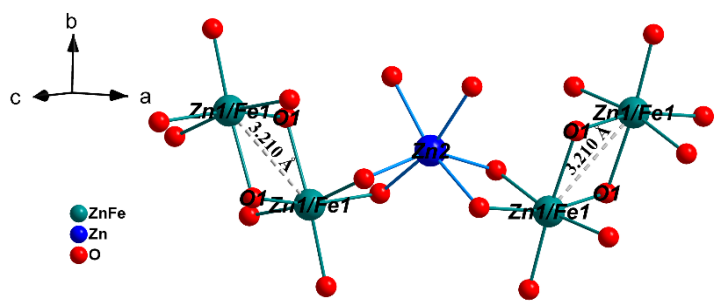

**Fig. 5** Representation of the super-exchange interactions between $Fe^{3+}$ in the dimers of $Na_2Zn_2Fe(VO_4)_3$

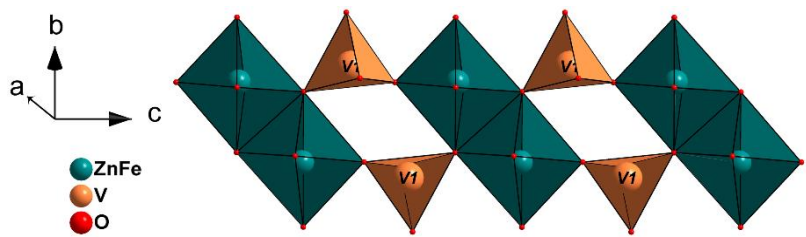

**Fig. 6** Representation showing the cation-anion-anion-cation super-exchange interactions path in $Na_2Zn_2Fe(VO_4)_3$

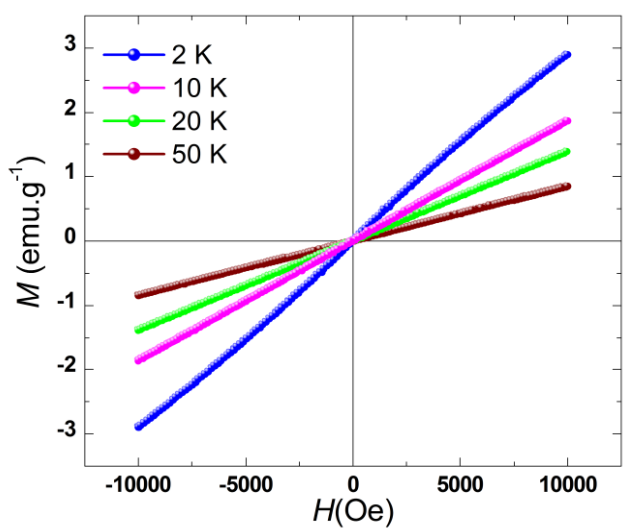

**Fig. 7** Isothermal magnetization *M-H* recorded between -10 kOe and 10 kOe at temperatures of 2, 10, 20, and 50 K

## Conclusion

In summary, single crystals and polycrystalline form of a novel alluaudite-type vanadate were successfully synthesized. The crystal structure was determined from single crystal X-ray diffraction data and powder X-ray diffraction confirmed the purity of the powder. Its crystal structure consist of a linkage of (Zn1/Fe1)$O_6$ octahedra, (Zn2)$O_6$ and $VO_4$ tetrahedra leading to an open 3D framework showing two type of wide tunnels where the $Na^+$ are located. This new vanadate was also characterized by scanning electron microscopy, Infrared spectroscopy and magnetic measurements. The magnetic investigation revealed a Curie-Weiss temperature of $\theta_{cw} = -42.36$ K suggesting antiferromagnetic interactions between $Fe^{3+}$ cations.

## Supplementary material

CSD 1920028 contains the supplementary crystallographic data for this paper. The data can be obtained free of charge from The Cambridge Crystallographic Data Centre via www.ccdc.cam.ac.uk/structures.

## Acknowledgements

This work was partially supported by CNRST (Centre National pour la Recherche Scientifique et Technique) in the Excellence Research Scholarships Program. This work was also supported by the European H2020-MC-RISE-ENGIMA action (n° 778072).

**Conflict of Interest:** The authors declare that they have no conflict of interest.